\documentclass[twocolumn,showpacs,aps,superscriptaddress]{revtex4}
\usepackage{graphicx}

\begin{document}

\draft
\title{Sling effect in collisions of water droplets in turbulent clouds}

\author{Gregory Falkovich$^1$ and Alain Pumir$^2$ }

\address{$^1$ Physics of Complex Systems, Weizmann Institute of Science, Rehovot 76100 Israel\\
$^2$ I.N.L.N., 1361 route des Lucioles, F-06560, Valbonne, France}



%

\begin{abstract}
We describe and evaluate the contribution of sling effect into the
collision rate of the same-size water droplets in turbulent
clouds. We show that already for Stokes numbers exceeding 0.2 the
sling effect gives a contribution comparable to Saffman-Turner
contribution, which may explain why the latter consistently
underestimates collision rate (even with the account of
preferential concentration).

\end{abstract}

\maketitle



The collision rate $N$ is the product of the target area $\pi
(a\!+\!a')^2$, the relative velocity of droplets before the
contact $\Delta v$ and the probability $P$ to find two droplets
touching:
\begin{equation}N(a,a')=\pi (a\!+\!a')^2\Delta vP(a+a')\ ,\end{equation}
where $a,a'$ are the radii of the droplets.  Here we focus on the
contributions of the air flow to the collision rate. To that end
we consider equal-size droplets which fall with the same velocity
in still air\footnote{There is a common misconception that the
gravitational collision rate is zero for equal-size droplets. This
is not so since hydrodynamic interaction changes settling
velocities (for example, a pair of close droplets falls faster).
For a dilute set of droplets, such effects can be neglected
though.} We also consider droplet sizes exceeding few microns and
neglect Brownian motion of droplets. For such droplets, the air
flow is the sole source of relative velocity $\Delta v$ and it
also influences $P(2a)$ due to droplet inertia.

Let us first briefly discuss the latter effect called preferential
concentration. It has been identified long ago  (see Maxey 1987;
Squires and Eaton 1991; Sundaram and Collins 1997; Reade and
Collins 2000; Kostinski Shaw 2001; Jaczewski and Malinowski 2005;
McFarquhar 2004; Franklin et al 2005; Grits et al 2006 and the
references therein). Still, the proper quantification of this
effect and of its role in the collision rate enhancement in warm
clouds remains to be done. It has been inferred from the data
(Sundaram and Collins, 1997) that $P(l)$ has a power-law
dependence and argued theoretically (Balkovsky et al, 2001) that
the dependence must be of the form $P(l)\sim (\eta/l)^\alpha$
where $\eta$ is of the order of the viscous scale of turbulence
and the dimensionless quantity $\alpha$ depends on the
dimensionless numbers that characterize air turbulence, particle
inertia and gravity --- Reynolds, Stokes and Froude numbers
respectively:
\begin{equation} Re=(L/\eta)^{4/3}\,,
\  St=\tau\nu/\eta^2=\lambda\tau\,,\
F=\lambda\eta/g\tau\,.\label{numbers}\end{equation} Here  $L$ is
the size of the largest turbulent eddies (typically, the size of
the cloud), $\nu$ is the air viscosity, $g$ is the acceleration of
gravity and $\tau\!=\!(2/9)(\rho_0/\rho)(a^2/\nu)$ is called the
reaction (Stokes) time with $\rho,\rho_0$ being the air and water
density respectively.

The behavior of $\alpha$ is well-understood for small $St$  where
$\alpha\simeq b(Re,F)St^2$ as predicted by Balkovsky et al (2001),
Falkovich et al (2002), and confirmed by Falkovich and Pumir
(2004) and Chu and Koch (2005) for moderate $Re$. Here we
establish $\alpha$ for arbitrary $St$ and substantially higher
$Re$ than before. Note in passing that a generalization for
different-size droplets value, $P(a,a')$, has been suggested in
Falkovich et al (2002) and Bec et al (2005).

The main subject of the paper is the proper evaluation of the
contribution of the relative velocity into the collision rate,
particularly, the (generally nonlocal) relation between the air
flow and the droplet velocities. Saffman and Turner (1956) assumed
that the relative velocity of the droplets is determined locally
by the air velocity which is spatially smooth at such scales
(since $a\ll\eta$) so that $\Delta v\simeq 2\lambda a$ where
$\lambda$ is the rms air-velocity gradient according to
(\ref{numbers}). However, the droplet velocity is determined not
only by a local air velocity but also by the previous history
because of inertia. As was first noticed by Falkovich et al
(2002), this leads to an extra contribution to the collision rate
not captured by the Saffman-Turner formula. Figure 1 illustrates
this (so-called sling) effect: The right droplet passed through an
intense vortex and had been thrown away as if by a sling. As a
result, the relative velocity of the droplets at the point of
collision may be determined not by the air-flow gradient at this
point but rather by a distant vortex. Let us give here some
numbers adopted from the calculations to be described below.
Consider turbulence with $\lambda=80\,s^{-1}$. For droplets with
$a=10\,\mu m$, $St=0.08$ and sling effects are negligible; for
$a=15\,\mu m$, $St=0.2$ and the frequency of sling effects is
$\simeq\lambda/1000$ while their contribution into the collision
rate is about $20\%$; for $a=20\,\mu m$, $St=0.35$ and the
frequency of sling effects is $\simeq\lambda/50$ while their
contribution into the collision rate is about $35\%$ (the numbers
correspond to Figs.~3,5,6 below). An independent confirmation that
Saffman-Turner formula (which disregards sling events)
consistently underestimates the collision rate even at relatively
small Stokes numbers has been obtained recently by direct
numerical simulations (Franklin et al, 2005).

\begin{figure}
\hskip 1truecm  \includegraphics[width=2in,angle=0]{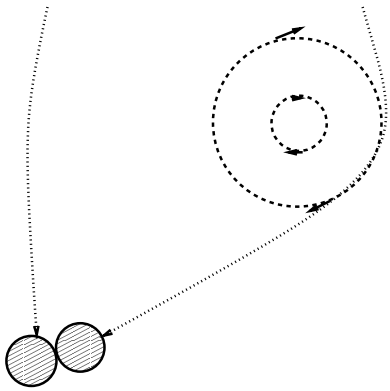}
  \caption{\label{fall}Sketch of the sling effect.
  Dotted lines show the trajectories of
the droplets while broken lines show the streamlines of the vortex
in the air flow. }
\end{figure}

The most straightforward way to model the collision rate is with
the discrete finite-size particles embedded into a properly
modelled flow (see, for instance Franklin et al, 2005 and Wang et
al, 2005). Since turbulent air flow and droplet distribution are
highly intermittent, the requirements on resolution and statistics
are such that both Reynolds numbers and number of droplets are
moderate at best in such a modelling. Moreover, computations with
discrete particles do not allow one to distinguish between local
contributions to collisions (that must be described by the
Saffman-Turner formula with a proper preferential concentration
correction) and a sling-effect contribution. That impedes the
progress towards a proper parametrization of the collision rate.
Here we propose another, complementary way of modelling based on
the continuous description of the flow of droplets.
 The equation for the droplet velocity
${\bf v}$,
\begin{equation}
d{\bf v}/dt= ({\bf u}-{\bf v})/\tau+{\bf
g}\,,\label{eqV}\end{equation} can be considered as defining
everywhere  in space the field ${\bf v}({\bf r})$ from the known
air-flow field ${\bf u}({\bf r})$. In the continuous description
we are able to see clearly the sling contribution. Indeed, sling
events appear in the equations of motion for particles as crossing
of trajectories or finite-time singularities of the continuous
equations. For the matrix of droplet-flow gradients,
$\sigma_{ij}=\partial v_i/\partial r_j$, one gets in the co-moving
(Lagrangian) reference frame:
\begin{equation}
\dot\sigma_{ij}+\sigma_{ik}\sigma_{kj}+\sigma_{ij}/\tau=s_{ij}/\tau\,,
\label{sigma1}\end{equation} where $s_{ij}=\partial u_i/\partial
r_j$. Smoothness of the air flow means finiteness of $s_{ij}(t)$.
On the contrary, nonlinearity (which corresponds to inertia)  in
the equation for $\hat\sigma$ leads to the possibility of
explosions when some component of $\sigma_{ij}$ turns into
$-\infty$ in a finite time by the law $\sigma\propto (t-t_0)^{-1}$
(Falkovich et al, 2002).

Because of spatial and temporal non-locality, analytical
description of the sling-effect contribution into the collision
rate is difficult. Using simple models, it has been predicted that
probabilities of such events have a very sharp dependence on the
Stokes number: $\exp[-c(Re,F)/St]$ (Falkovich et al 2002;
Wilkinson and Mehlig 2003,2005), where neither the value of the
factor $c(Re,F)$ nor its dependence on the Reynolds and Froude
numbers is known. Particularly important is understanding the
dependence on Reynolds number which varies by many orders of
magnitude in atmospheric flows.

In this paper, we perform direct numerical simulations of the
air-flow turbulence at such high $Re$ that were never reached
before in collision rate calculations. We pay the price by not
performing kinetic droplet simulations (i.e. following separate
droplets) but by using continuous field description based on
(\ref{eqV},\ref{sigma1}) and the equation for particle
concentration in the Lagrangian reference frame,
\begin{equation}
dn/dt=-n\sigma_{ii}(t)\ .\label{n}\end{equation} To this end, we
generate a statistically stationary turbulent flow in a cube with
periodic boundary conditions. The equations are solved with a
pseudo spectral code, see Pumir (1994) for details. All the
appropriate length scales of the flow are adequately resolved. We
work in the range $ 21 \le R_{\lambda } \le 105$. In this
turbulent flow, we follow the motion of inertial particles by
solving the equation for the position, ${ d {\bf x} / dt } = {\bf
v} $, along with (\ref{eqV}). The equation~ \ref{sigma1} for the
tensor of droplet velocity derivatives is integrated along the
way. The integration of Eq. ~\ref{eqV},\ref{sigma1} requires the
interpolation of the fluid velocity, ${\bf u}$, or its derivative,
$s$, from the numerical mesh to the particle position. This is
done by using spline interpolation techniques, see Girimagi and
Pope (1990). The equations are solved by using algorithms that are
second order accurate in time, or higher.

 For each run, we integrate the equations of motion for several Stokes numbers
(ranging from $St = 0.05$ to $St \approx 5$). Gravity is also
taken into account, by taking a finite value of the Froude number.
In this work, as in Falkovich and Pumir (2004), the only property
of the droplets which varies is the radius, $a$. Accordingly, the
Froude and the Stokes number vary in such a way that the product
of the Stokes number by the Froude number is constant :
\begin{equation}
\epsilon_0 \equiv  St \times Fr \label{epsilon0}
\end{equation}
Numerically, it was found that $\epsilon_0 \gtrsim 5$ corresponds to
a vanishingly
weak gravity. The effect of gravity becomes appreciable for values of
$\epsilon_0 $ smaller than $ 1$.

 The occurrence of a sling effect results in a singularity of $\sigma$ in a
finite time, $ \sigma \propto (t - t_0)^{-1}$, which in turn,
leads to a divergence of the particle density ($n \propto (t -
t_0)^{-1}$). Physically, neither the droplet velocity gradient nor
the droplet density can grow unrestricted since droplets have a
finite size, $a$, and cannot come arbitrarily close to one
another. Therefore, the droplet velocity gradients do not grow by
more than by a factor $\sim (\eta/a)$. Once the gradient has
reached this predetermined threshold value, the equation is
regularized, by flipping the sign of $\sigma$. This simply
corresponds to a fast droplet passing a slow one so that their
velocity difference changes sign (as well as the velocity gradient).
Similarly, the density increases until the time of the flip and
then decreases. This algorithm leads to a numerically well-posed
problem.

 To infer the properties of the coarse grained distribution at a given scale
$r$, we use the method developed by Balkovsky et al (2001),
Falkovich et al (2002) and implemented numerically by Falkovich
and Pumir (2004) in the restricted case where the Stokes number
was small, and where no sling contribution was expected. That
method requires one to follow (in addition to ${\bf x},{\bf
v},\sigma$ and $n$) the deformation of a volume, seeded with bubbles, and
carried by the ${\bf v}$-flow. Specifically, one needs to
determine the contraction rate of the volume along particle
trajectories. This is effectively done by monitoring the growth of
the inverse of the deformation tensor, $W$, which describes how a
line element is transported by the flow : $\delta {\bf l}(t) =
W(t) \cdot \delta {\bf l}(0)$, where $W^{-1}$ satisfies :
\begin{equation}
{ d W^{-1} \over dt } = - ( W^{-1} \cdot \sigma  + \sigma^T \cdot W^{-1} )
\label{Weq}
\end{equation}
To estimate the contribution of a trajectory to the coarse grained
particle density at scale $r$, the integration is carried out
until $W^{-1}$ reaches $(\eta/r)$. At this point, the value of $n$
is recorded. The Eulerian value of the $k^{th}$ moment is obtained
by averaging the values of $n^{k-1}$ over all the trajectories
computed.

In such an approach, the collision rate along a trajectory
consists of two contributions. The first one is the 'continuous
contribution', determined by a local velocity gradient (like in
the Saffman-Turner approach). Given the value of $n$ and of
$\sigma$ at a given time, the instantaneous flux of incoming
particles towards a given particle is as follows:
\begin{equation}
\Phi_{cont}(t) = - (2a)^3  n(t) \int_{\hat e \cdot \sigma \cdot \hat e < 0 }
( {\hat e \cdot \sigma \cdot \hat e } ) ~  d\Omega
\label{flux_cont}
\end{equation}
The contribution to the collision rate, $K_{cont}$, along a
trajectory is simply obtained by integrating $\Phi_{cont}$ over
time. Because the growth of $n$ is accounted for  in the resulting
contribution for the collision term, the influence of the sling
effect on concentration (because of caustics left after sling
events) is taken into account in this formula. However, as has
been pointed out above, the sling effect also results in an
additional contribution to the velocity difference, which is not
proportional to the local velocity gradient. As suggested by
Fig.1, a sling event involves a number of particles incoming in a
small region with significantly different velocities, a situation
leading to an outbursts of collisions, which we estimate as
follows. The source term in Eq.~\ref{sigma1}, necessary to start a
blow-up process, $s/\tau$, should be large enough to drive
$\sigma$ to start the blow up : $ |s| >
1/\tau$. Based on the Kolmogorov picture of fully developed
turbulence, the extent of the region where gradients reach the
value $s$ of the order or larger than $1/\tau$ is of the order of
$l \sim (\nu \tau)^{1/2} = \eta St^{1/2}$. The time over which the
collision takes place is estimated to be of the order of the
droplet relaxation time, $\tau$. Last, the range of particle
velocities involved during the collision can be estimated as $|
\delta {\bf v} | \sim l/\tau$. Based on these estimates, the
number of collisions that occur  in the wake of a sling event that
have happened at time $t_s$ can be estimated as follows:
\begin{equation}
N_{sling}(t) = 4 \pi (2a)^2 \times  n(t_s +\tau/2) \times | \delta {\bf v} |
\times \tau
\label{flux_sling}
\end{equation}
The relaxation time of droplet velocities is $\tau$ and this is a
typical duration of the time when droplets have substantially
different velocities and sling-effect contribution appears. That
is why in estimating $N_{sling}$ the density of particles is taken
at a time $\tau/2$ after the sling event time $t_s$, which
provides a reasonable estimate for the particle density during the
entire process. As it is clear from the derivation, the value of
$N_{sling}$ obtained in such a way is up to a numerical factor of
order unity, whose precise value could be estimated by kinetic
numerical simulations, which is beyond the scope of this work. In
the rest of this paper, we estimate separately the contributions
from the regular term and the sling contribution, given by
Eq.~\ref{flux_cont} and Eq.~\ref{flux_sling}, respectively.

 The method sketched above to estimate the coarse grained properties at a scale
$r$ is then used to evaluate the collision rate as a function of
scale. More precisely, consider droplets of radius $a$. To compute
the collision rate, the regular and sling contributions to the
collision rate are computed along trajectories, until the
compression, given by $| W^{-1} |$, see Eq.~\ref{Weq}, reach the
value $\eta/a$. The various contributions coming from different
trajectories are then accumulated, and the mean value of the
collision rate extracted.

 The way the flow leads to the compression of an ensemble of particles,
described by Eq.~\ref{Weq}, plays a crucial importance in the
physical processes controlling particle collision rates. The
numerical results indicate that $ln(|W^{-1}|)$ grows linearly with
time. The growth rate gives a direct access to the smallest
Lyapunov exponent $\lambda_3$ that corresponds to the strongest
contraction in the flow. At a given Reynolds number, the value of
$- \lambda_3 \times \tau_K $ has a non trivial dependence on the
Stokes number. It starts at a value $- \lambda_3 \times \tau_K
\approx 0.16 $ at $St = 0$ (Falkovich et al, 2002), then increases
up to $St \sim 0.5$, leading to a twofold increase of $\lambda_3$
compared to its value when $St = 0$, before decreasing at higher
values of $St$, see Fig.2. Qualitatively similar dependence has
been found for a model short-correlated flow (Wilkinson and Mehlig
2003). The values of $\lambda_3$ are not affected much by the
effect of gravity, at the moderate values of the Froude numbers
considered here, as shown in the figure. No significant dependence
of the product $\tau_K \times \lambda_3$ as a function of the
Reynolds number was found over the range covered here.

\begin{figure}
 \includegraphics[width=3.6in,angle=0]{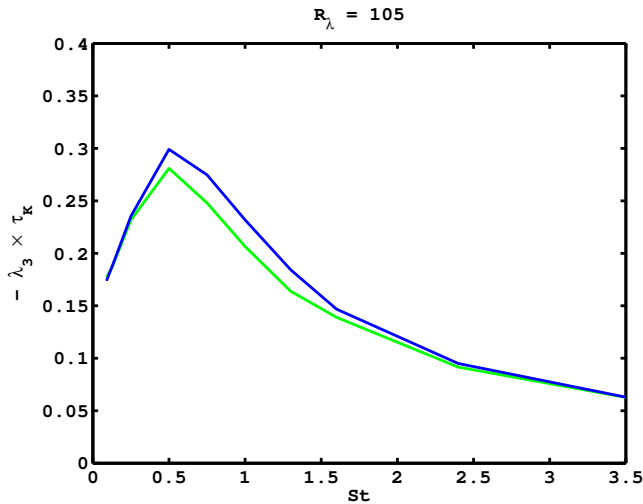}
 \caption{\label{lambda3} The contraction rate along trajectories of particles
at $R_{\lambda} = 105$ and at a very low gravity ($\epsilon_0 =
5$, upper curve) and at a moderate gravity ($\epsilon_0 = 0.4$,
lower curve). }

\end{figure}

Sling events, in our approach, are manifested by divergences of
the particle velocity derivative tensor, $\sigma$. The blow-up
frequency, $f_{bu}$, defined as the total number of sling events,
divided by the integration time, plays here a crucial role. Fig.~3
shows $f_{bu}$ multiplied by the Kolmogorov time $\tau_K$. The
data shown correspond to two values of $\epsilon_0$ ($\epsilon_0 =
5$ and $\epsilon_0 = 0.4$). The blow-up frequency, made
dimensionless with the Kolmogorov time, $\tau_K$, shows a fairly
weak dependence of the Reynolds number. No blow-up is observed at
very low value of $St$ (for $St \lesssim 0.15$). The value of
$f_{bu}$ then raises to a maximum value for $St \sim 1.5$, then
decreases slowly. Upon increasing gravity (decreasing
$\epsilon_0$), the blow-up frequency generally goes down, with a
similar Stokes number dependence. The recent works on simple
versions of the problem ($1$-dimensional and short-correlated
flows suggest the dependence of the blow-up frequency as a
function of $St$ of the form: $f_{bu} \approx exp(-A/St) $
(Wilkinson and Mehlig 2005, Derevyanko et al 2006). Here we find
empirically that the curve could be fit pretty well by the
dependence of the form:
\begin{equation}
f_{bu} \times \tau_K = St^{-2} \times exp(-A/St) \times (B + C
St^{c}) \label{fit}
\end{equation}
The coefficient $A$ is found to decrease slightly as the Reynolds number
increases ($A = 2.1$ for $R_\lambda = 45$, $A=1.85$ for $R_\lambda = 83$
and $A = 1.70$ for $R_\lambda = 105$),
consistent with the fact that as turbulence becomes more intense,
higher gradients appear in the flow, which are able to induce blow-up of
$\sigma$ at increasingly low values of $St$. Yet, at higher values of
the Stokes number, the blow-up frequency seems to {\it decrease} as the
value of the Reynolds number increases, a somewhat surprising effect.

\begin{figure}
 \includegraphics[width=3.3in,angle=0]{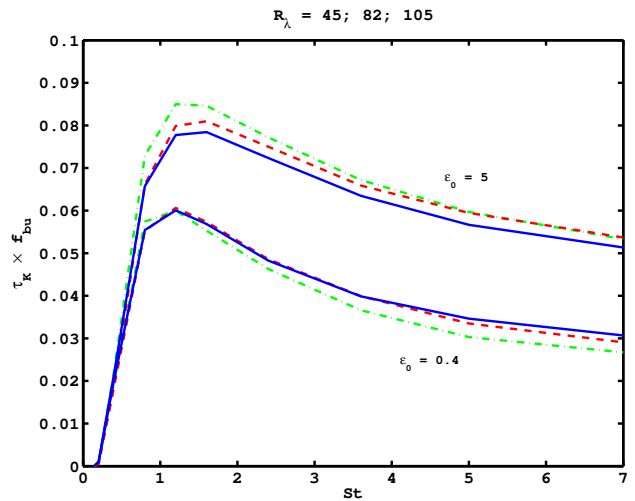}
 \caption{\label{blow_up_frequency}The blow up frequency as a function of the
Stokes numbers for several Reynolds numbers: $R_\lambda = 45$
(dot-dashed line), $83$ (dashed line) and $105$ (full line) and at
$\epsilon_0 = 5$ and $\epsilon_0 = 0.4$. }

\end{figure}

 The dependence of the coarse-grained particle density, $\langle n^2
\rangle_r$ as a function of $\eta/r$, is very similar to the one
obtained by Falkovich and Pumir, 2004. Namely, $\langle n^2
\rangle_r$ has essentially a power-law dependence as a function of
$\eta/r$. The exponent $\alpha$ of the exponent is plotted here as
a function of $St$ at $R_{\lambda} = 83 $, and for the values of
$\epsilon_0 = 5$ (very low gravity) and $\epsilon_0 = 0.4$
(moderate gravity). The value of the exponent increases sharply as
a function of $St$ up to $St \approx 1$, where it starts to
saturate and decrease slightly. The qualitative aspect of the
dependence of $\alpha$ as function of $St$ does not depend on the
precise value of the Reynolds number in the range of $R_\lambda$
studied. We find that at values of $St \ge 0.1$, the value of the
exponent is somewhat higher than the one found by Falkovich et al,
2002, see Fig.~4. This difference can be attributed to the fact
that the exponent there was computed by studying contraction along
the {\it fluid} trajectory, which in the limit $St \rightarrow 0$
differs very little from the {\it particles} trajectories.
Quantitative differences remain even for values of $St$ as small
as $St \approx 0.1$. At moderate values of $St$, the exponent is
larger when gravity is small, as expected (Falkovich et al 2002).
At larger values of the Stokes number ($St \gtrsim 2 $), the two
exponents obtained with a very low gravity ($\epsilon_0 = 5$) and
with a moderate gravity ($\epsilon_0 = 0.4$) become very close to
one another.

\begin{figure}
 \includegraphics[width=3.6in,angle=0]{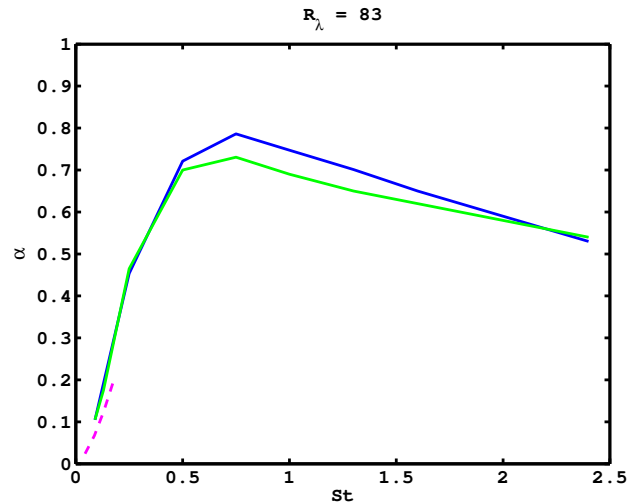}
\caption{ \label{alpha_vs_St} The $St$-dependence of the exponents
$\alpha$, obtained by fitting $ \langle n^2 \rangle_r$ by a power
law dependence as a function of $r$ : $ \langle n^2 \rangle_r
\propto (\eta/r)^{\alpha}$, at a very small value of gravity
($\epsilon_0 = 5$, upper curve) and at a moderate value of gravity
($\epsilon_0 = 0.4$, lower curve). The value of the exponent
obtained by Falkovich and Pumir (2004) at small values of $St$, is
shown by the dashed line. }
\end{figure}


 Finally, Fig.~5 shows the continuous contribution to the collision rates,
normalized by $32\pi a^3\tau_K$, as a function of the Stokes
numbers. The continuous contribution starts at a low value at $St
\rightarrow 0$, close to the value predicted by the Saffman and
Turner formula (indicated by the horizontal dashed line), and
increases to a maximum value at $St \approx 1.$, before decreasing
very slowly. Increasing gravity (decreasing $\epsilon_0$) tends to
decrease the collision rate. Over the range of parameters studied
here, it was found that the continuous part of the collision rate
increases when the Reynolds number increases.

\begin{figure}
 \includegraphics[width=3.2in,angle=0]{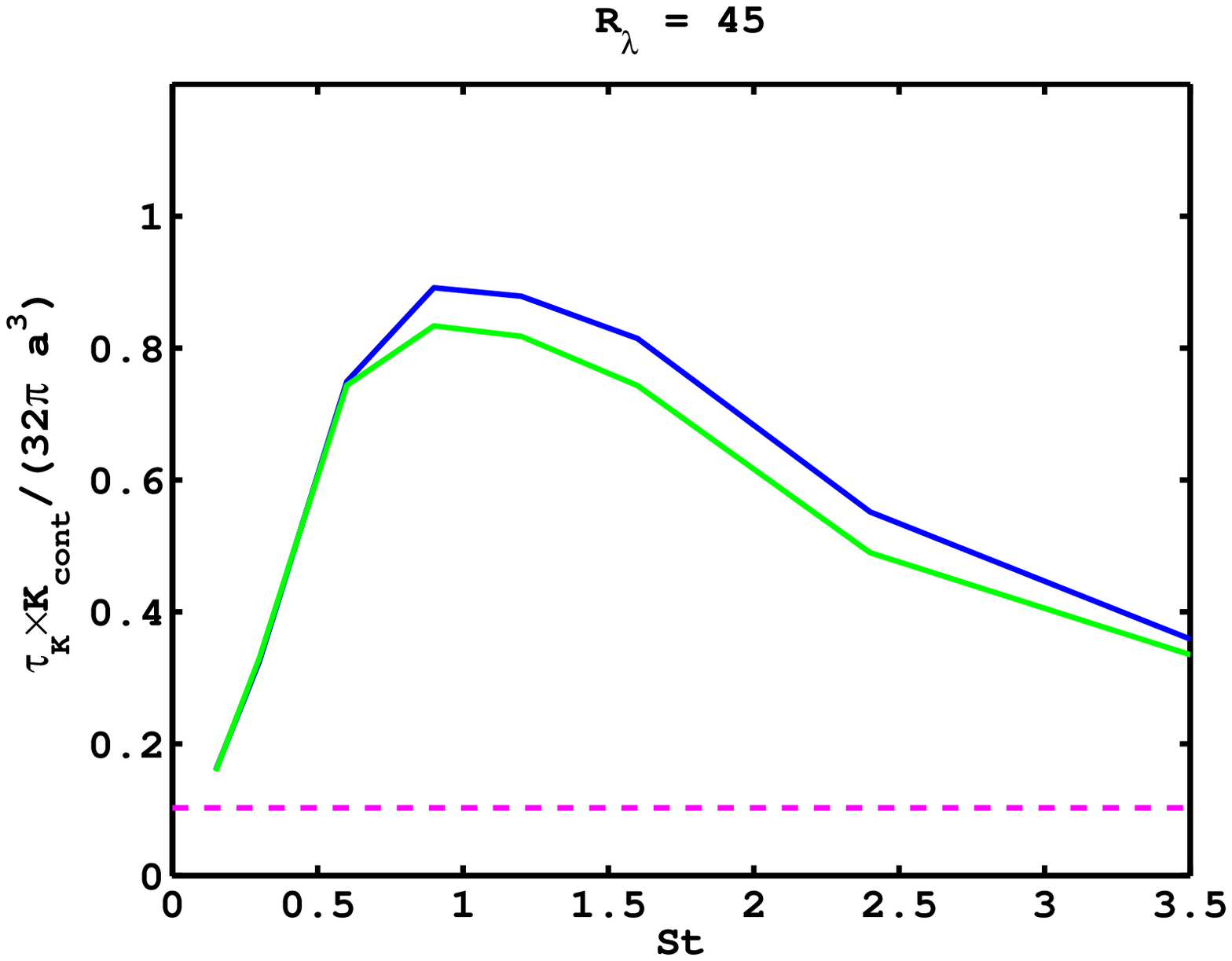}
 \includegraphics[width=3.2in,angle=0]{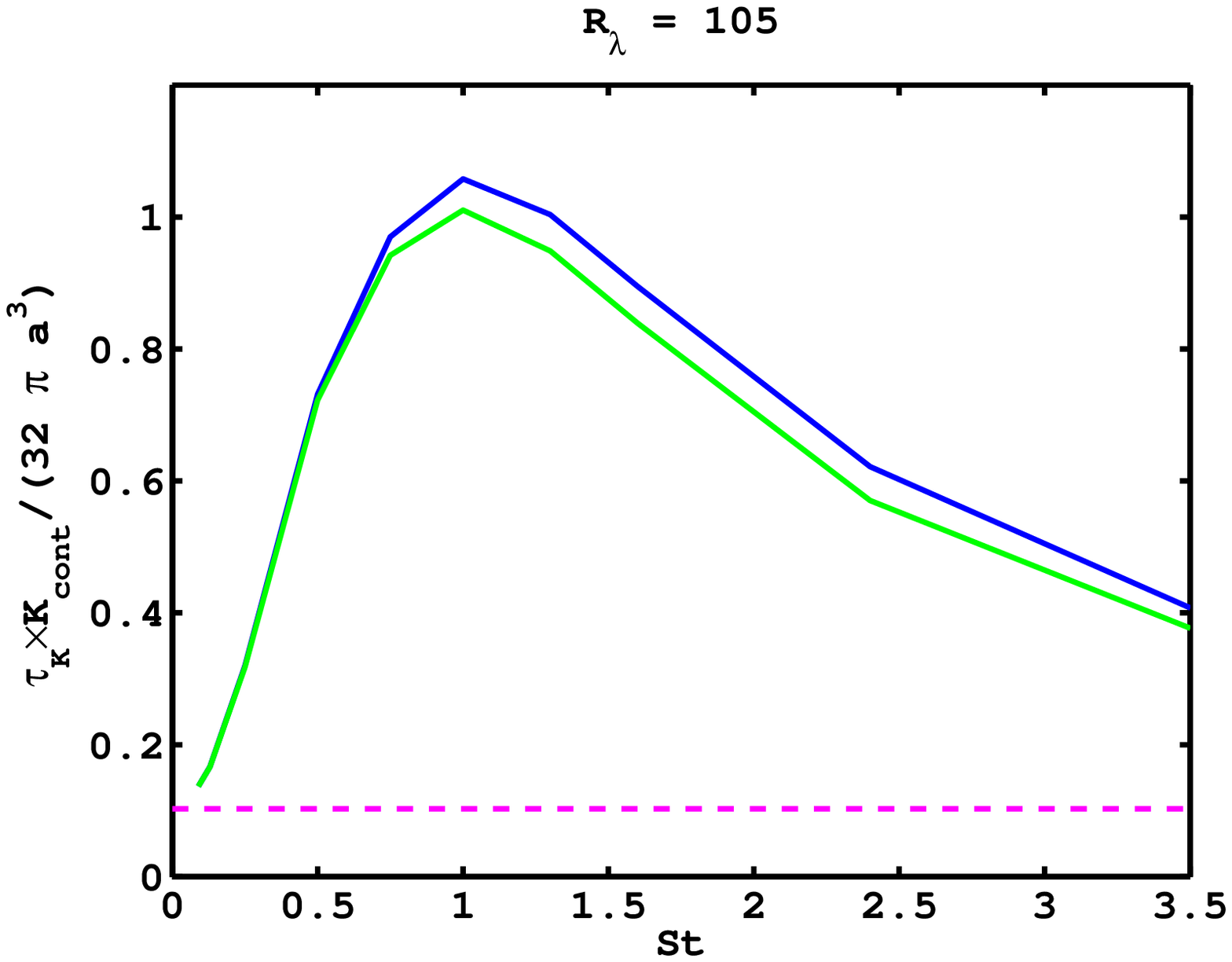}
\caption{ \label{Kcont} The $St$-dependence of the
continuous component of the collision rate, normalized by the particle
size, $a$ and by the
Kolmogorov time scale, $\tau_K$, at small ($\epsilon_0 = 5$, upper curve)
and moderate ($\epsilon_0 = 0.4$, lower curves) value of gravity.
The horizontal dashed line corresponds to the Saffman and Turner formula.
The upper graph corresponds to a Reynolds number of $R_{\lambda} = 45$, the
lower graph to a Reynolds number of $R_{\lambda} = 105$.
The collision rate peaks at a value $St \approx 1$. }
\end{figure}

 The sling contribution to the collision rate, shown in Fig.~6 at the value of
the Reynolds number $R_\lambda = 105$, starts from essentially
zero at very small values of the Stokes number (the probability of
having a sling effect is practically zero at $ St \ll 1$). Again,
similarly to what has been observed for the continuous
contribution to the collision term, it increases to a maximum at
$St \sim 0.8$. The phenomenological description of the sling
collision rate used in this work is not expected to hold at values
of the Stokes numbers larger than $\sim 1$.  In fact, our approach
is based on the implicit assumption that particles can be
described by an essentially smooth hydrodynamic representation.
This assumption becomes questionable as soon as $St \gtrsim 1$.
For this reason, only the part of the curve corresponding to
values of $St \le 1.0 $ has been shown. Fig.5 is meant to show the
main trend, at moderate Stokes numbers (the formula used to define
this term, Eq.\ref{flux_sling} is defined up to a constant).
 These data do
not allow to test the simple approximation $St^{-1/2}exp(-A/St)$
for the sling contribution suggested by Wilkinson et al (2006) for
the case without gravity.

\begin{figure}
 \includegraphics[width=3.2in,angle=0]{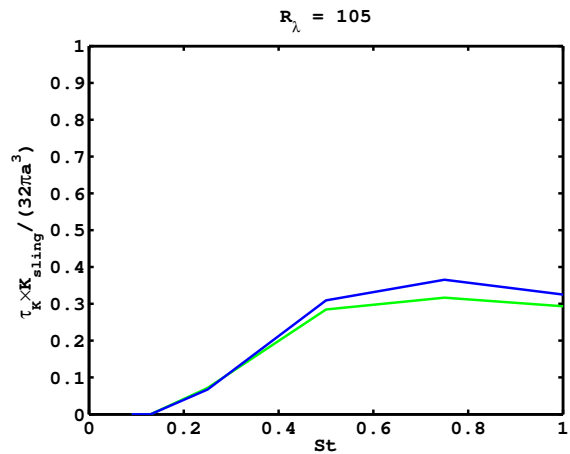}
\caption{ \label{Ksling} The $St$-dependence of the
sling component of the collision rate, normalized by the particle
size, $a$, and by the Kolmogorov time, $\tau_K$,
at small ($\epsilon_0 = 5$, upper curve)
and moderate ($\epsilon_0 = 0.4$, lower curves) value of gravity, at
a Reynolds number of $R_{\lambda} = 105$.
}
\end{figure}

 Our method to estimate collision rates, although based on procedures which
can be formally completely justified, is very indirect. It is
therefore appropriate to compare our estimates for the collision
rates with the results obtained by Franklin et al., 2005, by using
direct numerical simulations, and by estimating in a
straightforward manner the collision rate among particles. We ran
a simulation at a value of $R_\lambda = 45$ (run 3a) comparable to
the run 3 of Franklin et al, at $R_\lambda = 48$. The values of
their 'Geometric collision rates', estimated for two particles'
sizes : $a = 10 \mu m$ ($\Gamma_{10}$) and $a = 20 \mu m$
($\Gamma_{20}$), see their Table 5, are to be compared directly to
the value of $K_{cont}$ and of $ K_{sling}$, computed in the
present work. The values obtained with our method, after proper
rescaling, are shown in Table \ref{comparison_table}. At the
lowest value of $St$, the value of $\Gamma$ and $K_{cont}$ are
very close. Part of the difference can be explained by the fact
that our Reynolds is slightly less than the one obtained by
Franklin et al. No sling contribution is expected in this case. At
the highest value of the Stokes number, one finds that the
continuous part of the collision rate, $K_{cont}$, underestimates
the value found in Franklin et al. On the other hand, a
significant sling effect is expected at this value of the Stokes
number, so the difference can be   interpreted as resulting from
the sling contribution.

\begin{table*}
\caption{Comparison between the numerical estimates of Franklin et al., 2005,
and the present estimates. The subscript $10$ (respectively $20$)
 refer to particles of size $10 \mu m$ (respectively $20 \mu m$).}
\label{comparison_table}
\begin{center}
\begin{tabular}{l c c c c c c c  }
\hline
Run \# & $R_{\lambda}$ & $a$ & $St$ &
$\epsilon_0$ & $\Gamma$ & $K_{cont}$ & $K_{sling}$ \\
\hline\hline
3  & $48$ & $10 \mu m$ & $0.08$ &
$0.21$ & $1.0\times10^{-6}$ &  &  \\
3a  & $45$ &  & 0.08 &
$0.2$  & $0$ & $8.5\times10^{-7}$  & $0$ \\
\hline
3  & $48$ & $20 \mu m$ & $0.32$ &
$0.21$ & $5.9\times10^{-6}$ &  &  \\
3a  & $45$ &  &  &
$0.2$ &  & $2.2 \times 10^{-6}$  & $1 \times 10^{-6} $ \\
\hline\hline
\end{tabular}
\end{center}
\end{table*}

  In conclusion, we have studied the collision rates induced by
turbulent air motion. The method used in this work is essentially
lagrangian. We follow particles advected in the flow, compute
directly the flux of incoming particles (continuous contribution)
and estimate the number of collisions occuring in the aftermath of
a 'sling' effect.

 The ratio of the collision rate to the Saffman-Turner
formula is found to increase significantly  from $1$  to $\sim 10$
when the Stokes number increases from $St \approx 0$ to $St
\approx 1$.  The increase becomes more pronounced as the Reynolds
number becomes larger.

 In the range of Reynolds number studied here, sling contributions are
negligible at very small Stokes numbers: their probability goes as
$\exp(-A/St)$ as a function of $St$, with a coefficient $A $ of
order $1$. In practice, they become significant for Stokes numbers
$St \gtrsim 0.20$.

 The actual collision rates computed in this work are consistent,
at $R_{\lambda} \approx 45$, with the recent results obtained by Franklin
et al. (2005). In particular, our results allow us to disentangle the
contributions due to the sling events, which we find to be quite significant
for particles of size $a = 20 \mu m$.

 This work should help clarify the origin of the enhancement of the collision
rates of inertial particles due to turbulence, and also ultimately,
to devise a parametrization of this collision rate.

%

We acknowledge support by the Israel Science Foundations and by the European
Commission (Contract HPRN-CT-2002-00300) and by IDRIS for computer time.
G.F. is grateful to A. Kostinski for useful discussions.


\bibliography{mybibfile}

\end{document}